\documentclass[aps,prc,reprint,amsmath,amssymb,showpacs,preprintnumbers,superscriptaddress,nofootinbib]{revtex4-1}
\usepackage{iftex}
\ifPDFTeX
  \usepackage{CJKutf8}
\else
  \usepackage{xeCJK}
\fi
\usepackage{graphicx}
\usepackage{subcaption}
\usepackage{dcolumn}
\usepackage{bm}
\usepackage{color}
\usepackage{hyperref}

\allowdisplaybreaks[4]

\begin{document}

\ifPDFTeX
\begin{CJK*}{UTF8}{gbsn}
\fi
\title{
Efficient emulation of nuclear ground states with neural-network variational Monte Carlo and eigenvector continuation
}
\author{Mao Li (李茂)}
\affiliation{State Key Laboratory of Nuclear Physics and Technology, School of Physics, Peking University, Beijing 100871, China}

\author{Yilong Yang (杨一龙)}
\affiliation{State Key Laboratory of Nuclear Physics and Technology, School of Physics, Peking University, Beijing 100871, China}

\author{Pengwei Zhao (赵鹏巍)}
\email{pwzhao@pku.edu.cn}
\affiliation{State Key Laboratory of Nuclear Physics and Technology, School of Physics, Peking University, Beijing 100871, China}

\begin{abstract}
An efficient emulator for \emph{ab initio} calculations of nuclear ground-state properties is developed by integrating the neural-network variational Monte Carlo framework, \emph{FeynmanNet}, with the eigenvector continuation.
It enables the calculation of observables for different Hamiltonians with minimal computational cost, while delivering ground-state energies with errors below $0.5\%$ compared to the full \emph{FeynmanNet} results.
With this emulator, the ground-state energies and charge radii of ${}^{16}\mathrm{O}$, ${}^{15}\mathrm{O}$, ${}^{14}\mathrm{O}$, ${}^{15}\mathrm{N}$, and ${}^{14}\mathrm{C}$ are computed using a nuclear Hamiltonian derived from the leading-order pionless effective field theory, with a large number of different values of low-energy constants (LECs).
Then, we perform a global sensitivity analysis of the ground-state energies, charge radii, separation energies of selected nuclei for the three LECs in the Hamiltonian, to identify how each LEC contributes to the variances of these observables.
It shows that the two-body LEC in the $^3S_1$ channel is the most influential LEC governing these nuclear bulk properties.
Finally, the correlations among the ground-state energies of $^4$He, $^{12}$C, and $^{16}$O are investigated by varying the LECs in the Hamiltonian.
The analysis reveals that the experimental ground-state energies of $^{12}$C and $^{16}$O cannot be reproduced simultaneously by varying the LECs in the leading-order pionless Hamiltonian.
This suggests that additional ingredients in the leading-order Hamiltonian are required to improve its description of light nuclei.
The present work establishes an efficient framework for global sensitivity analysis and uncertainty quantification in the quantum Monte Carlo calculations for light and medium-mass nuclei. 
\end{abstract}

\maketitle
\ifPDFTeX
\end{CJK*}
\fi

\section{Introduction}
Understanding the emergent properties of atomic nuclei from the interaction of protons and neutrons is a longstanding goal in nuclear theory.
Many models of two-nucleon (NN) interactions have been constructed by high-quality fits to NN scattering observables up to high energies, either phenomenologically~\cite{Wiringa1995Phys.Rev.C38,Machleidt2001Phys.Rev.C024001} or in the framework of chiral effective field theory (EFT)~\cite{Gezerlis2014Phys.Rev.C054323,Piarulli2016Phys.Rev.C054007,Entem2017Phys.Rev.C024004,Reinert2018Eur.Phys.J.A86}.
These high-resolution interactions provide crucial input for accurate \emph{ab initio} many-body calculations, enabling significant advances in our understanding of light nuclei.
In particular, the continuum quantum Monte Carlo (QMC) calculations based on the Argonne $v_{18}$~\cite{Wiringa1995Phys.Rev.C38} NN interaction plus the Urbana/Illinois three-nucleon (3N) interactions~\cite{Pudliner1997Phys.Rev.C1720,Pieper2001Phys.Rev.C014001} are able to reproduce many low-lying states, moments, and transitions in light nuclei up to $A=12$.
Recently, many QMC studies of light nuclei have been conducted using local chiral NN+3N interactions~\cite{Gezerlis2014Phys.Rev.C054323, Piarulli2016Phys.Rev.C054007}, which provide systematically improvable theoretical descriptions of nuclear structure~\cite{Lonardoni2018Phys.Rev.Lett.122502,Piarulli2018Phys.Rev.Lett.052503}, electroweak~\cite{King2020Phys.Rev.C025501,Martin2023Phys.Rev.CL031304,ChambersWall2024Phys.Rev.Lett.212501}, and scattering properties~\cite{Lynn2016Phys.Rev.Lett.062501,Flores2023Phys.Rev.C034001} of light nuclei.
However, such high-resolution interactions feature substantial high-momentum components and sophisticated spin-isospin structures.
They incur severe sign problems that exponentially exacerbate with increasing number of nucleons, posing significant challenges on accurate QMC calculations beyond light nuclei.

Although high-resolution interactions are essential for describing observables that probe high-momentum dynamics, such as lepton-scattering processes~\cite{Chen2017Phys.Rev.Lett.262502,CruzTorres2021NaturePhysics306310}, simplified low-resolution interactions might already be sufficient for describing many low-momentum nuclear bulk properties.
Based on such tenet, the quest for the ``essential elements of nuclear binding'' was proposed~\cite{Lu2019Phys.Lett.B134863} to seek the simplest nuclear Hamiltonian for reproducing the nuclear bulk properties.
In Ref.~\cite{Lu2019Phys.Lett.B134863}, a lattice nuclear Hamiltonian derived within leading-order (LO) pionless EFT was constructed for lattice QMC calculations, able to yield binding energies and radii for $A\leq50$ nuclei with a few percent errors.
Thanks to the simplicity of the LO pionless EFT Hamiltonian, it enables sign-problem-free QMC calculations of many-nucleon systems on the lattice~\cite{Shen2021Eur.Phys.J.A276,Lu2022Phys.Rev.Lett.242501,Ren2024Phys.Lett.B138463,Tong2025ApJ164}.
Recently, the wave function matching method~\cite{elhatisari2024Wavefunction} has also been shown to drastically mitigate the sign problem in lattice QMC calculations.
For continuum QMC calculations, an essential Hamiltonian ``o'' inspired by LO pionless EFT was also constructed, and it can reproduce reasonably well the binding energies of several light and medium-mass closed-shell nuclei~\cite{Schiavilla2021}.
These phenomenological successes are supported by the arguments based on the unitary limit, large $N_c$, and SU(4) symmetry~\cite{Koenig2017Phys.Rev.Lett.202501, Kievsky2018Phys.Rev.Lett.072701,Lu2019Phys.Lett.B134863,Lee2021Phys.Rev.Lett.062501}.

Recently, the progress in seeking the essential elements of nuclear binding has been significantly advanced by the development of variational Monte Carlo (VMC) approaches based on neural-network wave functions.
This approach makes use of the ability of artificial neural networks to compactly represent many-body wave functions~\cite{Carleo2017}.
It enables accurate variational solutions to the quantum many-body problem with polynomial scaling and inherently circumvents the sign problem.
Based on LO pionless EFT Hamiltonians, the Slater-Jastrow wave functions were developed for neural-network VMC calculations of $A\leq 6$ nuclear ground states~\cite{Adams2021,Yang2022Phys.Lett.B137587,Gnech2021FewBodySystems7}.
Later, the incorrect nodal surface of the Slater-Jastrow wave functions was addressed by augmented Slater determinants with hidden nucleons~\cite{Lovato2022}, which enabled accurate solutions of the $^4$He and $^{16}$O ground states.
Alternatively, a neural-network architecture \emph{FeynmanNet}~\cite{Yang2023} employed the spin-isospin dependent backflow transformation to improve the wave functions' nodal surface, providing accurate solutions of $^4$He, $^6$Li, and $^{16}$O ground states.
Recently, the neural-network Pfaffian-Jastrow wave function~\cite{Fore2025Commun.Phys.108}, following its success in ultra-cold Fermi gases~\cite{Kim2024Commun.Phys.148}, was developed to calculate dilute nuclear matter by simulating up to $A=42$ nucleons in periodic boxes.
Using the essential Hamiltonian ``o''~\cite{Schiavilla2021}, the neural-network-based VMC calculations are able to yield a neutron matter equation of state that is remarkably close to high-resolution Hamiltonians at low densities~\cite{Fore2023}.
In addition, such calculations describe well the experimental data on the dipole response of $A\leq4$ nuclei~\cite{Parnes2025} and the magnetic moments of nuclei up to $A=20$~\cite{Gnech2024Phys.Rev.Lett.142501}. 

Apart from the simplified pionless EFT Hamiltonians, neural-network wave functions suitable for solving high-resolution Hamiltonians have also been developed.
In Ref.~\cite{Yang2025ChinesePhys.Lett.051201},  the ground states of $A\leq 4$ nuclei were solved using the high-precision Bonn potential~\cite{Machleidt1989189376}, with 
the ``Jastrow" wave function consisting of two-body correlation functions represented by neural networks and two-body spin-isospin operators.
Subsequently, backflow transformation is implemented in the correlation functions of the Jastrow wave function, facilitating the solution of $A=5$ neutron-$^4$He scattering problems~\cite{yang2025chiralsymmetry}.
Recently, Ref.~\cite{yang2025zemach} proposed a novel functional form of neural-network wave functions, which enables virtually-exact variational solutions of $A\leq 7$ nuclei with high-resolution Hamiltonian.

Given the success of the essential Hamiltonian and the accuracy of neural-network VMC approaches, it is crucial to study how the predictions of the observables depend on its model parameters, i.e., the low-energy constants (LECs) in the pionless EFT, by performing the global sensitivity analysis (GSA)~\cite{Sobol2001,Ekstrom2019,Sun2025}.
In addition, to enable a comparison between the theoretical predictions and experimental data in rigorous statistical terms, the employment of uncertainty quantification with Bayesian approaches is useful~\cite{Furnstahl2015JPG034028}.
In turn, GSA and uncertainty quantification can provide guidance on the best way to extract the LECs from experimental data. 
They require solving the nuclear many-body problem with different values of LECs in the Hamiltonian; however, repeating the full \emph{ab initio} calculations multiple times can become computationally prohibitive.
An alternative approach is to substitute the full \textit{ab initio} calculation with a computationally less expensive emulator~\cite{Drischler2022}, such as the eigenvector continuation (EC)~\cite{Frame2018,Melendez2022,Duguet2024}.
EC is a reduced-basis method that builds a low-dimensional variational subspace from snapshot eigen-wave-functions at selected parameter values, and then uses that subspace to interpolate or extrapolate eigenvalues and eigenfunctions for other model parameters.
The convergence properties of EC have been analyzed in Ref.~\cite{Sarkar2021}.
The method has been widely applied to nuclear ground states~\cite{Ekstrom2019,Koenig2020,demol2020Improved,wesolowski2021Rigorous,djarv2022Bayesian,Yapa2022,companysfranzke2024Eigenvector}, excited states~\cite{yoshida2022Constructing,companysfranzke2022Excited,becker2023Initio}, nuclear matter~\cite{jiang2024Emulating}, nuclear scattering and reactions~\cite{Furnstahl2020,Melendez2021,Drischler2021,bai2021Generalizing,Liu2024,Lei2026}, and nuclear resonances~\cite{Yapa2023,yapa2025Scalablea}.
Given the high quality of neural-network wave functions, it is appealing to combine neural-network VMC calculations and the EC approach.

In this work, we integrate the neural-network VMC approach based on \emph{FeynmanNet} architecture with the eigenvector continuation to build an efficient emulator, dubbed \emph{FeynmanNet}-EC, for sensitivity studies and uncertainty quantification.
The accuracy of the \emph{FeynmanNet}-EC emulator is benchmarked against the full \emph{FeynmanNet} calculation.
As a first application, we perform a global sensitivity analysis~\cite{Sobol2001} of the binding energies, separation energies, and charge radii of several $A\simeq 16$ nuclei for the LECs in the essential Hamiltonian inspired by pionless EFT~\cite{Schiavilla2021}.
This allows the identification of the most influential LECs in the essential Hamiltonian that governs nuclear bulk properties.
Finally, we apply the \emph{FeynmanNet}-EC emulator to study the correlations among the $^4$He, $^{12}$C and $^{16}$O ground-state energies by varying the LECs, to provide insights into the alpha-alpha interactions as emerged from the essential Hamiltonian. 

\section{Theoretical framework}
\subsection{Eigenvector continuation with \emph{FeynmanNet}}
Eigenvector continuation is based on the observation that, under smooth variations of some real-valued control parameters in the Hamiltonian, trajectories of an eigenvector lie on a low-dimensional manifold within the high-dimensional Hilbert space~\cite{Frame2018}.
Therefore, the eigenvectors on the trajectory can be well approximated by linear combinations of the eigenvectors at a few parameter values.
Specifically, we consider nuclear Hamiltonians derived within effective field theories (EFTs), varying smoothly with the low-energy constants (LECs) $\bm\alpha$, 
\begin{equation}\label{eq:H}
    H(\bm \alpha)=H_0+\sum_{n} \alpha_n V_n,
\end{equation}
where $H_0$ is the part independent of the LECs, including the kinetic and Coulomb energy, and $V_n$ are the short-range potentials.
The ground-state eigenvector $|\Psi^\star\rangle$ at the target LECs $\bm \alpha^\star$ is expanded on the ground-state eigenvectors $\{|\Psi^{(i)}\rangle\}_{i=1,2\ldots,N_{\rm EC}}$ at $N_{\rm EC}$  training points, $\{\alpha^{(i)}\}_{i=1,2\ldots,N_{\rm EC}}$.
Then, the best approximation to $|\Psi^\star\rangle$ is obtained from the following generalized eigenvalue problem,
\begin{equation}
H(\bm{\alpha}^\star) u = \lambda Nu,
\label{eq:EC}
\end{equation}
with the norm matrix $N$ and the projected Hamiltonian matrix $H$,
\begin{align}
    N_{ij}=\frac{\langle\Psi^{(i)}|\Psi^{(j)}\rangle}{\sqrt{\langle\Psi^{(i)}|\Psi^{(i)}\rangle \langle\Psi^{(j)}|\Psi^{(j)}\rangle}},\\
    H_{ij}=\frac{\langle\Psi^{(i)}|H|\Psi^{(j)}\rangle}{\sqrt{\langle\Psi^{(i)}|\Psi^{(i)}\rangle\langle\Psi^{(j)}|\Psi^{(j)}\rangle}}.
\end{align}
The lowest eigenvalue $\lambda$ approximates the ground-state energy at LEC value $\bm\alpha^\star$ and the $u_i$ are the expansion coefficients,
\begin{equation}
    |\Psi^\star\rangle=\sum_{i=1}^{N_{\rm EC}}u_i\frac{|\Psi^{(i)} \rangle}{\sqrt{\langle \Psi^{(i)}|\Psi^{(i)}\rangle}}.
\label{eq:wavefunction}
\end{equation}
The full nuclear many-body problem is now reduced to a straightforward solution of the $N_{\rm EC}\times N_{\rm EC}$-dimensional generalized eigenvalue problem.
The computational cost of the latter is negligible compared to that of the former, enabling rapid emulations of the ground-state properties for a large number of LEC sets.

Apart from the ground-state energy, EC also supports the reconstruction of the emulated eigenvector and, thus, the emulation of other ground-state observables.
In this work, we select the root-mean-square (rms) charge radius as an example of such an observable for investigation.
For the ground state of the target LECs $\bm \alpha^\star$ 
we have obtained in Eq.~(\ref{eq:wavefunction}) , we can express the mean-square charge radius as

\begin{equation}
  \langle r^2_{\rm{ch}} \rangle= \langle r^2_{\rm{pt}} \rangle + \langle R^2_p \rangle + \frac{N}{Z} \langle R^2_n \rangle + \frac{3}{4M_p^2},
\end{equation}
where $\langle r^2_{\rm{pt}} \rangle$ is the point-proton mean-square radius, $\langle R^2_p \rangle$ and $\langle R^2_n \rangle$ are the mean-square charge radii of the proton and neutron, respectively~\cite{Beringer2012}, and the last term is the Darwin-Foldy relativistic correction~\cite{Friar1997} with $M_p$ the proton mass.
The point-proton mean-square radius is defined as
\begin{equation}
    \begin{split}
        \langle r^2_{\rm{pt}} \rangle & = \frac{1}{Z}\langle \Psi^\star | \sum_i P_p \bm r_i^2 | \Psi^\star \rangle \\
        & = \sum_{i,j=1}^{N_{\rm EC}} u_i u_j \frac{\langle \Psi^{(i)} | \sum_{a=1}^AP_{p,a} \bm r_a^2 | \Psi^{(j)} \rangle}{\sqrt{\langle \Psi^{(i)}|\Psi^{(i)} \rangle \langle \Psi^{(j)}|\Psi^{(j)} \rangle }},
    \end{split}
\end{equation}
with $P_{p,a}=(1-\tau_{za})/2$ the proton projection operator of the $a$th nucleon. 
Then, we can compute the contributions from all the $(i,j)$ elements in advance and store them for subsequent calculations varying LECs.
For any LEC set, we only need to calculate the coefficients $u_i$ through Eq.~(\ref{eq:EC}) to get the point-proton mean-square radius~\cite{Gnech2021FewBodySystems7}.

In the present calculations, the norm matrix $N$ and the projected Hamiltonian matrix $H$ are estimated via Monte Carlo sampling in the coordinate space.
First, $N_{\rm EC}$ sets of samples are respectively drawn from the distributions of $N_{\rm EC}$ eigen wave functions,
$\bm X^{(i)}\sim |\Psi^{(i)}|^2$ with $\bm X$ including the spatial, spin, and isospin coordinates of all nucleons.
Then, the matrix elements of $N$ and $H$ are estimated by the expectation values on the prepared samples.
It is crucial to calculate the matrix elements with the same sets of Monte Carlo samples, rather than redrawing samples individually for each matrix element.
In this way, we can take advantage of the statistical correlations between the matrix elements to reduce the statistical fluctuations of the outcome of the generalized eigenvalue problem in Eq.~(\ref{eq:EC}).
In addition, for the Hamiltonian matrix, we calculate each term in Eq.~(\ref{eq:H}) individually,
\begin{eqnarray}
    H_{ij}=(H_0)_{ij}+\sum_{n}\alpha_n (V_n)_{ij},
\end{eqnarray}
so that the matrix elements for $H_0$ and $V_n$ need to be evaluated only once and used for all the subsequent calculations varying the LECs.
This significantly reduces the computational cost.

For the norm matrix $N$ and Hamiltonian matrix $H$, their $(i,j)$ matrix elements are evaluated on both the two sets of samples $\bm X^{(i)}\sim |\Psi^{(i)}|^2$ and $\bm X^{(j)}\sim |\Psi^{(j)}|^2$ by taking the geometric mean of the two evaluations,
\begin{align}\label{eq.MC}
    O_{ij}=e^{i\varphi}\sqrt{\frac{\langle \Psi^{(i)} | O|\Psi^{(j)} \rangle}{\sqrt{\langle \Psi^{(i)} | \Psi^{(i)} \rangle\langle \Psi^{(j)} | \Psi^{(j)} \rangle}} \cdot \frac{\langle \Psi^{(j)} |O| \Psi^{(i)} \rangle}{\sqrt{\langle \Psi^{(i)} | \Psi^{(i)} \rangle \langle \Psi^{(j)} | \Psi^{(j)} \rangle}}},
\end{align}
where 
\begin{align}
\frac{\langle \Psi^{(i)} | O| \Psi^{(j)} \rangle}{\sqrt{\langle \Psi^{(i)} | \Psi^{(i)} \rangle \langle \Psi^{(j)} | \Psi^{(j)} \rangle}}   = \frac{\left| \left\langle \frac{O\Psi^{(j)}(\bm X^{(i)})}{\Psi^{(i)}(\bm X^{(i)})}\right\rangle\right| }{\sqrt{\left\langle\left|\frac{\Psi^{(j)}(\bm X^{(i)})}{\Psi^{(i)}(\bm X^{(i)})}\right|^2\right\rangle }} \cdot e^{i \varphi},
\end{align}
\begin{align}
\frac{\langle \Psi^{(j)} | O| \Psi^{(i)} \rangle}{\sqrt{\langle \Psi^{(i)} | \Psi^{(i)} \rangle \langle \Psi^{(j)} | \Psi^{(j)} \rangle}}   = \frac{\left|\left\langle \frac{O\Psi^{(i)}(\bm X^{(j)})}{\Psi^{(j)}(\bm X^{(j)})}\right\rangle\right| }{\sqrt{\left\langle\left|\frac{\Psi^{(i)}(\bm X^{(j)})}{\Psi^{(j)}(\bm X^{(j)})}\right|^2\right\rangle }} \cdot e^{-i \varphi},
\end{align}

with $\langle \cdot\rangle$ denoting the expectation values on the Monte Carlo samples.
The operator $O$ is unity for calculating $N$, and it is $H_0$ or $V_n$ for calculating the terms in the Hamiltonian.
The phase of the matrix element $\varphi$ can be taken from either of the two evaluations.
By computing with the two wave function distributions separately and then taking their geometric mean, we ensure that the 
$H$ and $N$ matrices respect the exchange symmetry of the $i,j$ indices, thereby preserving their Hermiticity.

Training the EC emulator requires accurate eigenvectors, in the present case, ground-state wave functions for a set of $N_{\rm EC}$ points across the chosen parameter domain of the LECs.
In this work, the ground-state wave functions at the training points are computed with the neural-network VMC framework, \emph{FeynmanNet}~\cite{Yang2023}.
Given a training point $\bm\alpha$, the wave function is parameterized by a neural-network ansatz of the following functional form,
\begin{equation}\label{Eq:Psi}
  \Psi(\bm x_1,\ldots,\bm x_A)=\mathrm{e}^{\mathcal{U}(\bm x_1,\ldots,\bm x_A)}\sum_{n=1}^{N_\mathrm{det}}\det[\mathbf{f}^{(n)}(\bm x_1,\ldots,\bm x_A)],
\end{equation}
with $\bm x_i=(\bm r_i,s_i,t_i)$ the single-nucleon variables including the spatial, spin, and isospin coordinates.
It is a product of permutation-invariant Jastrow correlator $\mathcal{U}$ and a sum of $N_{\rm det}$ Slater determinants consisting of backflow transformed orbitals $\mathbf{f}^{(n)}$.
All the $\mathcal{U}$ and $\mathbf{f}^{(n)}$s are represented by neural networks with specific architectures designed to respect the antisymmetry of the nuclear wave function; see Ref.~\cite{Yang2023} for their detailed expressions.
Moreover, the parity (time-reversal symmetry for even-even nuclei) is preserved explicitly by multiplying Eq.~(\ref{Eq:Psi}) with $1+\pi \hat{P}$ ($1+\hat{T}$) with $\hat{P}$ the space inversion operator ($\hat{T}$ the time-reversal operator).
Throughout the present calculations, the size of neural networks is taken to be the same as in Ref.~\cite{Yang2023} and the number of determinants is taken to be $N_{\rm det}=4$.

The parameters in \emph{FeynmanNet} are optimized by exploiting the variational principle,
\begin{equation}
E[\Psi] = \frac{\langle \Psi | \hat{H}(\boldsymbol{\alpha}) | \Psi \rangle}{\langle \Psi | \Psi \rangle}.
\label{eq:energy}
\end{equation}
Starting from a trial wave function $\Psi$, the energy expectation is minimized iteratively using the stochastic reconfiguration algorithm~\cite{Sorella2005Phys.Rev.B241103}, and the ground-state wave function is obtained after the energy minimization reaches a plateau~\cite{Yang2023}.
Instead of randomly initializing the neural-network parameters, we use the so-called transfer learning technique~\cite{Gnech2024Phys.Rev.Lett.142501} to efficiently train the neural networks using different LECs.
To be specific, we use the neural-network parameters trained for the original LEC set~\cite{Schiavilla2021} as the initial point for the training with varied LECs.

\subsection{Nuclear Hamiltonian}
We employ the essential Hamiltonian (model ``o") in Ref.~\cite{Schiavilla2021}, inspired by pionless EFT at leading order.
It is based on the tenet that the typical momenta of nucleons in nuclei are much smaller than the pion mass~\cite{Hammer2020Rev.Mod.Phys.025004}.
Therefore, the low-energy nuclear properties can be approximately described by a simplified Hamiltonian containing only regularized contact interactions.
In particular, the essential Hamiltonian adopted in this work takes the form
\begin{equation}\label{eq.HLO}
H = \sum_{i=1}^A \frac{-\nabla_i^2}{2M}+\sum_{i<j} v_{ij}^C
+ \sum_{i<j} v_{ij}
+ \sum_{i<j<k} V_{ijk},
\end{equation}
including the nonrelativistic kinetic energy with $M$ the nucleon mass, the Coulomb potential between finite-size protons~\cite{Wiringa1995Phys.Rev.C38}, and the regularized contact NN and 3N interactions, $v_{ij}$ and $V_{ijk}$.

The NN interaction reads
\begin{equation}
v_{ij} =C_{01} 
\frac{
e^{-r_{ij}^2/R_1^2}}{\pi^{3/2} R_1^3}  P^{\sigma}_0 P^{\tau}_1
+ 
C_{10}\frac{e^{-r_{ij}^2/R_0^2}}{\pi^{3/2} R_0^3} 
 P^{\sigma}_1 P^{\tau}_0 ,
\end{equation}
where $r_{ij}=|\bm r_i-\bm r_j|$ and the spin and isospin projection operators read
\begin{equation}
P^{\sigma}_0 = \frac{1 - \boldsymbol{\sigma}_1 \cdot \boldsymbol{\sigma}_2}{4}, \quad
P^{\sigma}_1 = \frac{3 + \boldsymbol{\sigma}_1 \cdot \boldsymbol{\sigma}_2}{4},
\end{equation}
with analogous definitions for $P^{\tau}_0$ and $P^{\tau}_1$ for isospin.
The LEC $C_{01}$ ($C_{10}$) and radii $R_1$ ($R_0$) control the strength and range of $T=1$ ($T=0$) NN interactions, respectively.
In this work, the model ``o" of Ref.~\cite{Schiavilla2021} is employed, whose cutoff radii $R_0=1.54592984$ fm and $R_1=1.83039397$ fm and LEC values $C_{01}=-5.27518671$ fm$^2$ and $C_{10}=-7.04040080$ fm$^2$ are determined by fitting to the neutron-proton scattering lengths and effective ranges in the $S$-wave singlet and triplet channels, and the deuteron binding energy. 

The 3N interaction takes the regulated Gaussian form
\begin{equation}\label{eq:3NF}
V_{ijk} = 
\frac{c_E}{f_{\pi}^4 \Lambda_\chi} 
\frac{1}{\pi^3 R_3^6}
\sum_{\mathrm{cyc}}e^{-{(r_{ij}^2+r_{ik}^2)}/{R_3^2}},
\end{equation}
where $\Lambda_\chi=1$ GeV, $f_\pi=92.4$ MeV is the pion decay constant, and the sum runs over cyclic permutations of the nucleon indices.
The optimal set of $c_E=1.0786$ and $R_3=1.0$ fm is adopted, which yields a reasonable description of the binding energies of several closed-shell nuclei up to $^{90}$Zr~\cite{Schiavilla2021} and the magnetic moments of $A\leq 20$~\cite{Gnech2024Phys.Rev.Lett.142501}.
In the following sensitivity analysis, the above mentioned values of LECs are taken as reference values.

\subsection{Global sensitivity analysis}
The essential Hamiltonian is governed by the three LECs, $C_{01}$, $C_{10}$, and $c_E$. 
These couplings characterize the strengths of the two-body spin–isospin channels and the three-body force, respectively. 
Following Ref.~\cite{Ekstrom2019}, we perform a global sensitivity analysis (GSA) to analyze how the predictions of nuclear ground-state properties depend on these LECs.
GSA provides a systematic means to learn how much each model parameter contributes to the uncertainty in a model prediction~\cite{Sobol2001}, and has been successfully applied to high-resolution chiral Hamiltonians~\cite{Ekstrom2019,Sun2025,Belley2024,belley2025}.
It differs from an uncertainty analysis, which addresses the question of how uncertain the prediction itself is. 
However, GSA is very computationally demanding and requires a tremendous number of sample calculations.
The integration of \emph{FeynmanNet} with EC effectively addresses GSA's computational intensity, enabling efficient evaluation of model outputs for given parameter sets with minimal computational overhead.
After the emulator is trained, approximately 2000 calculations with varied LECs can be performed in one second on a NVIDIA A800 80GB platform.

Let $Y$ denote an observable (e.g., ground-state energy or rms charge radius) viewed as a function of the LEC vector $\boldsymbol{\alpha}$. The Sobol variance decomposition expresses the total variance as~\cite{Sobol2001}
\begin{equation}
\mathrm{Var}[Y] = \sum_{i} W_i + \sum_{i<j} W_{ij} + \cdots ,
\label{eq:vardec}
\end{equation}
with first-order and second-order partial variances
\begin{align}
&W_i = \mathrm{Var}\left[\, \mathbb{E}_{\vec{\alpha}\sim (\alpha_i)}[Y \mid \alpha_i] \,\right],\\
&W_{ij}=\mathrm{Var}\left[\, \mathbb{E}_{\vec{\alpha}\sim (\alpha_i,\alpha_j)}[Y \mid \alpha_i,\alpha_j]\right]-W_i-W_j \,,
\label{eq:Vi}
\end{align}
where $\mathrm{Var}\left[\, \mathbb{E}_{\vec{\alpha}\sim (\alpha_i)}[Y \mid \alpha_i]\,\right]$ 
denotes the variance of the conditional expectation of $Y$, and $\vec{\alpha}\sim (\alpha_i)$ denotes the set of all
LECs excluding $\alpha_i$.

The corresponding sensitivity indices read
\begin{equation}
S_i = \frac{W_i}{\mathrm{Var}[Y]},\,\,S_{ij} = \frac{W_{ij}}{\mathrm{Var}[Y]}.
\label{eq:Si}
\end{equation}
The total-effect index $S_T(\alpha_i)$ accounts for all contributions involving $\alpha_i$, including interactions,
\begin{equation}
S_T(\alpha_i) = S_i + \sum_{j\neq i} S_{ij} + \cdots .
\label{eq:ST}
\end{equation}
The first-order sensitivity, $S_i$, is often referred to as the main effect.
It apportions the total variance in the model output to an individual model parameter $\alpha_i$. 
The higher-order indices apportion the variance in the model output to the interaction of parameters. 
To compute the Sobol sensitivity indices, we employ the Saltelli sampling method~\cite{Saltelli1999} to generate samples with different parameters.
First, one constructs two independent $N\times k$ matrices $A$ and $B$, each of which consists of a total of $N$ random samples of parameter values ($k$ is the number of parameters).
Then, by replacing one column of $A$ (or $B$) with the corresponding column of $B$ (or $A$), one can construct an additional matrix with different samples; repeating such an operation for each column of $A$ and $B$ generates a total of $2k$ different sampled matrices.
Finally, one evaluates the model output for all the samples in the above $2k$ matrices, along with $A$ and $B$, yielding a total of $N(2k+2)$ sampled model outputs to compute the Sobol sensitivity indices.
Such a procedure enables the stochastic evaluation of how model outputs change when only a single parameter value is varied, which is essential for estimating the Sobol sensitivity indices.
In this work, we set $N=1024$ and the number of LECs $k=3$, resulting in a total of $N(2k+2)=8192$ sampled model outputs.

For each sample, the corresponding observable is evaluated via \emph{FeynmanNet}-EC emulator, which enables Monte Carlo estimation of Sobol sensitivity indices with negligible cost per sample.
To evaluate the final result and statistical uncertainty and calculate the mean and the standard deviation, we repeatedly calculate each Sobol sensitivity index by 100 times with different random seeds.

\section{Numerical details}
The training points in the LEC parameter space must be properly chosen to achieve accurate EC emulation.
In particular, the EC subspace should be constructed from a representative set of 
wave functions such that the span of these states adequately covers the relevant manifold of eigenvectors.
  
\begin{figure}
  \centering
  \includegraphics[width=1\linewidth]{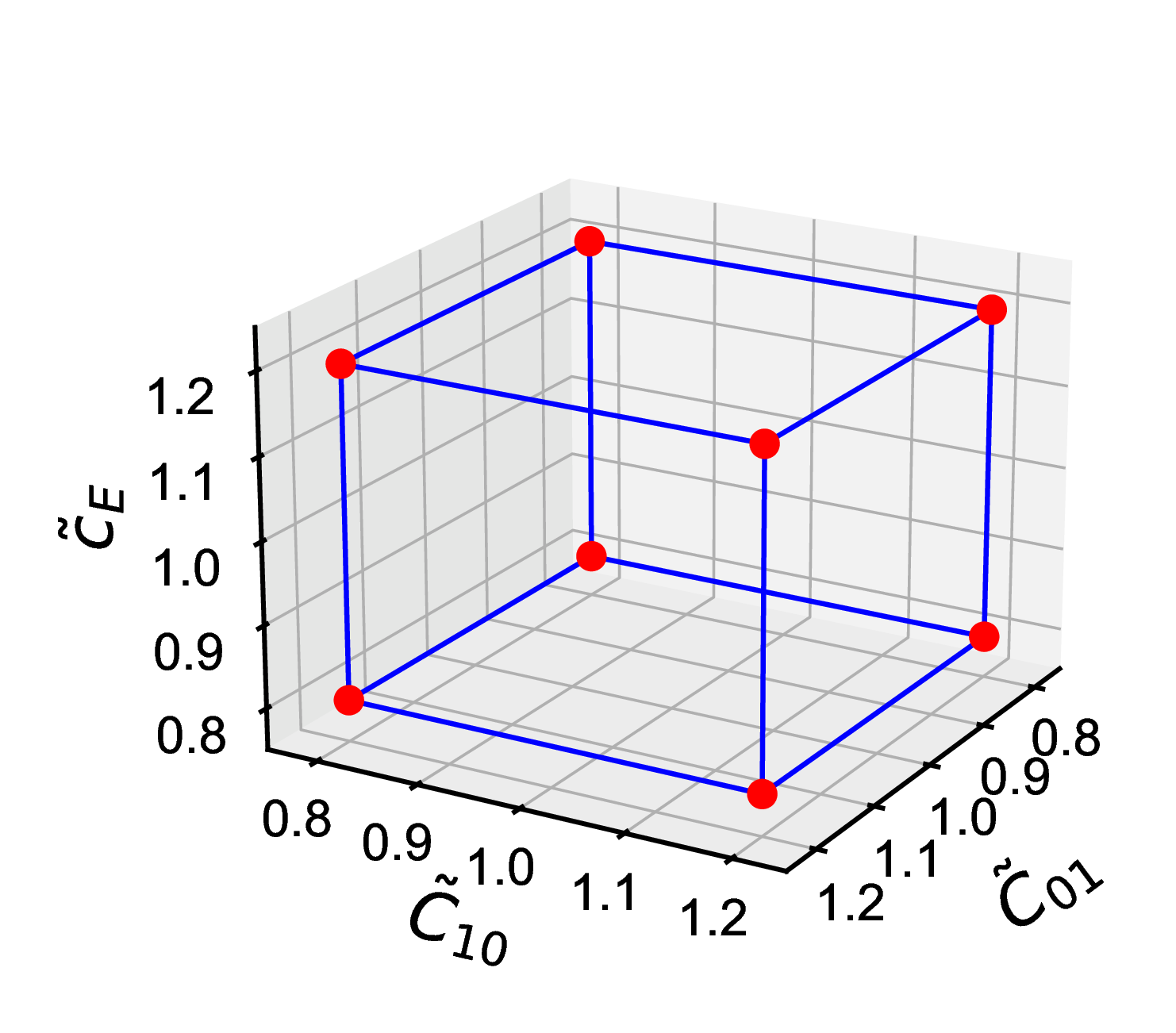}
  \caption{Distribution of the selected EC samples in the three-dimensional 
  parameter space spanned by the low-energy constants 
  \(C_{01}\), \(C_{10}\), and \(c_E\). 
  $\tilde{C}_{10}$, $\tilde{C}_{01}$, and $\tilde{c}_{E}$ denote the ratio of the LECs to their reference values determined in Ref.~\cite{Schiavilla2021}.
  The eight cube vertices correspond to 
  \(\pm 20\%\) variations of each LEC around its reference value, defining the training set for 
  constructing the EC subspace.
}
  \label{fig:lec_samples}
\end{figure}

In this work, we select the EC samples by varying the three LECs 
(\(C_{01}\), \(C_{10}\), and \(c_E\)) within a cubic region of parameter space, 
defined by simultaneous \(\pm 20\%\) deviations around their reference values determined in Ref.~\cite{Schiavilla2021}. 
The eight vertices of this cube serve as the $N_{\rm EC}=8$ sample points used to generate \emph{FeynmanNet} wave functions for EC construction (see Fig.~\ref{fig:lec_samples}). 
For the subsequent sensitivity analysis, the LECs are sampled from a uniform distribution within this cube.

The EC emulation relies on Monte Carlo evaluations of the norm and Hamiltonian matrix.
To estimate the statistical uncertainties of the EC emulation, we employ the bootstrap resampling method~\cite{Efron1993} as a non-parametric tool for estimating error bars.
For each EC emulation, we create multiple new bootstrap datasets from the original Monte Carlo sample dataset. 
Each dataset is generated by randomly sampling, with replacement, from the original dataset to form a new dataset of the same size. 
We create a large number of new datasets, each of them yields a bootstrap estimate of the EC-emulated observable, and the standard deviation across all such estimates is taken as the statistical uncertainty.

In addition, to perform stable EC emulation, it is crucial to avoid the numerical singularities in the norm matrix $N$. 
Unless $N_{\rm EC}$ is very small, in practice, the EC subspace can easily contain vectors that are almost linearly dependent, leading to a nearly singular norm matrix $N$.
The small eigenvalues of $N$ could turn to zero or negative due to statistical fluctuations in the Monte Carlo evaluations, resulting in an ill-conditioned generalized eigenvalue problem  [Eq.~\eqref{eq:EC}].
To address this issue, we employ singular value decomposition (SVD) to 
truncate the eigenmodes of $N$ whose eigenvalues fall below a chosen 
threshold $\varepsilon$ ~\cite{Hansen1987}.
First, we diagonalize the norm matrix,
\begin{equation}
N = U \, \Lambda \, U^{\dagger}, 
\end{equation}
where $\Lambda = \mathrm{diag}(\lambda_1, \lambda_2, \dots, \lambda_{N_{\rm EC}})$ 
contains the eigenvalues (singular values) of $N$ and $U$ is the corresponding unitary matrix. 
We then introduce a cutoff $\varepsilon$ and retain only those 
eigenmodes with $\lambda_i > \varepsilon$. 
Let $\tilde{U}$ denote the submatrix of $U$ corresponding to the retained eigenvectors, 
and $\tilde{\Lambda}$ the diagonal matrix of the retained eigenvalues.
The truncated transformation matrix is then
\begin{equation}
M = \tilde{U} \, \tilde{\Lambda}^{-1/2}.
\end{equation}

In the reduced $k$-dimensional subspace spanned by the retained modes, 
the eigenvalue problem takes the form
\begin{equation}
\tilde{H} \, \tilde{u} = \lambda  \tilde{u},
\end{equation}
with the reduced $k\times k$ Hamiltonian matrix $\tilde{H} = M^{\dagger} H M$ and the $k$-dimensional eigenvector $\tilde{u}=M u$.

\begin{figure}[!htbp]
  \centering
  \includegraphics[width=1\linewidth]{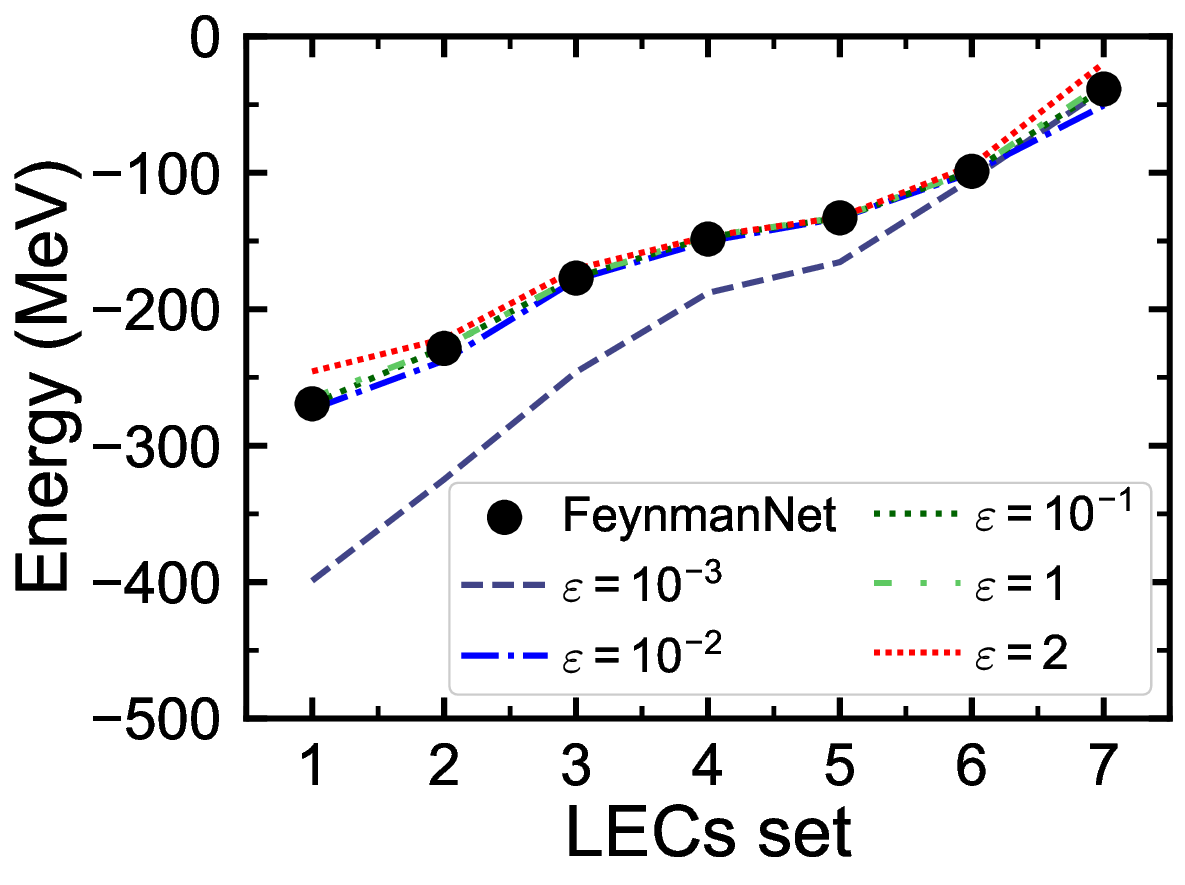}
  \caption{Comparison of the EC emulation for ${}^{16}\mathrm{O}$ ground-state energy across seven sampled LEC points, under different truncation thresholds $\varepsilon$ applied to the eigenmodes of the norm matrix $N$. }
  \label{fig:svd_truncation}
\end{figure}

\begin{figure}[!htbp]
  \centering
  \includegraphics[width=1\linewidth]{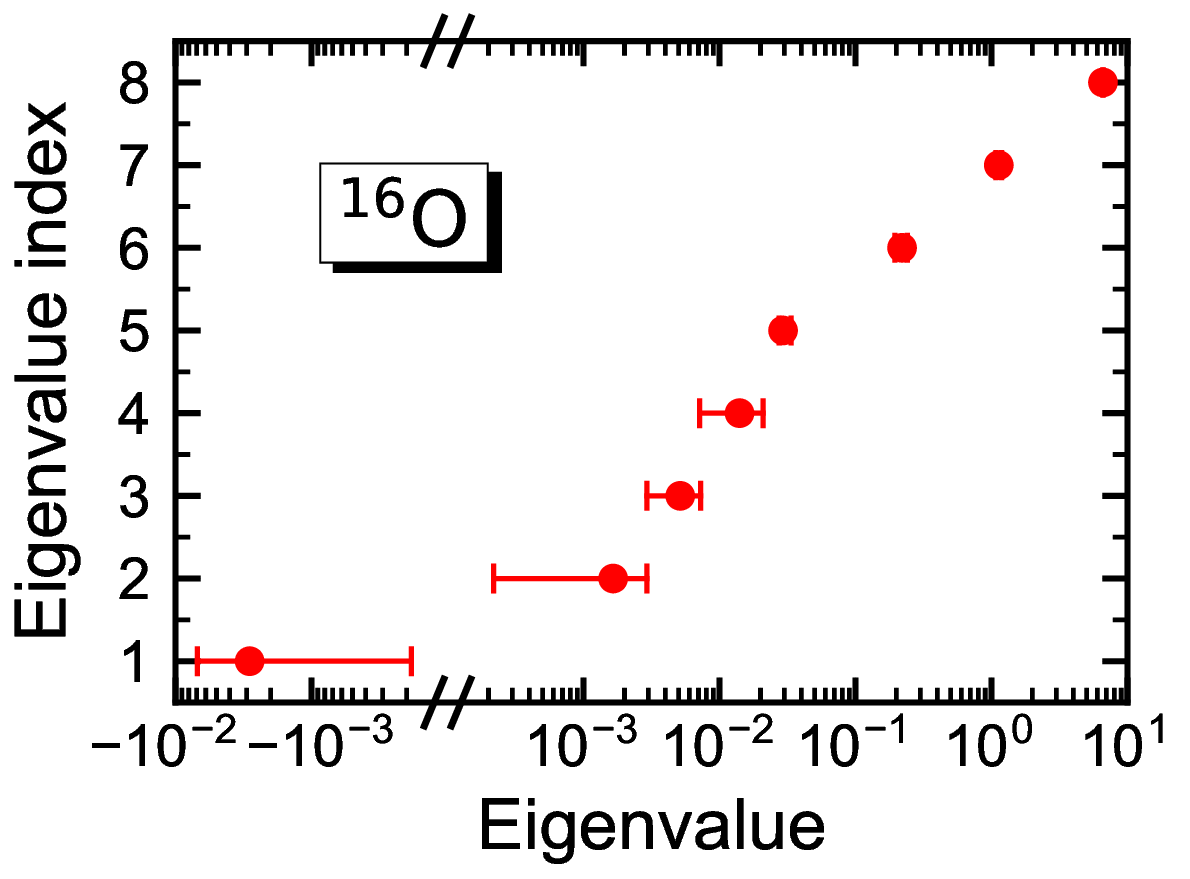}
  \caption{The eigenvalues of the norm matrix with $N_{\rm EC} = 8$ for $^{16}\mathrm{O}$. For each eigenvalue, the red dot and the error bar represent its mean and range, respectively, estimated from six sets of statistically independent Monte Carlo samples. }
  \label{fig:N_spectrum}
\end{figure}

Figure~\ref{fig:svd_truncation} illustrates how the truncation of singular values of $N$ stabilizes the EC emulation, taking the ground-state energy of $^{16}$O as an example.
The EC emulated results obtained under different truncation thresholds $\varepsilon$ are compared to the direct \emph{FeynmanNet} results for several randomly drawn LEC points.
In the case $\varepsilon=10^{-3}$, i.e., without truncation, the EC results lead to pronounced deviations from \emph{FeynmanNet}.
In contrast, the truncation with $\varepsilon=10^{-2}$-$1$ yields EC results consistent with \emph{FeynmanNet}.
For higher truncation thresholds, $\varepsilon\approx 2$, the performance of EC deteriorates, due to disregarding useful information in the emulation.

Figure~\ref{fig:N_spectrum} depicts the eigenvalue spectrum of the norm matrix $N$. The appearance of negative eigenvalues indicates that the Monte Carlo sampling noise should be the main limiting bottleneck of the present emulation, as it has already induced unphysical components in the norm matrix. Therefore, the number of EC training points stops at $N_{\rm EC}=8$ in the present work. The eigenvalues below $10^{-2}$ exhibit error bars comparable to their magnitudes. Retaining these eigenmodes may destabilize the generalized eigenvalue problem, justifying the choice of $\varepsilon=10^{-2}$ as a proper truncation threshold.
Therefore, in realistic applications, an appropriate eigenvalue truncation strategy guided by the spectrum of $N$ is essential to achieve reliable and precise EC predictions.
Throughout this work, we adopt $\varepsilon=10^{-2}$.

\section{Results and discussion}

\begin{figure}[!htbp]
  \centering
  \includegraphics[width=1\linewidth]{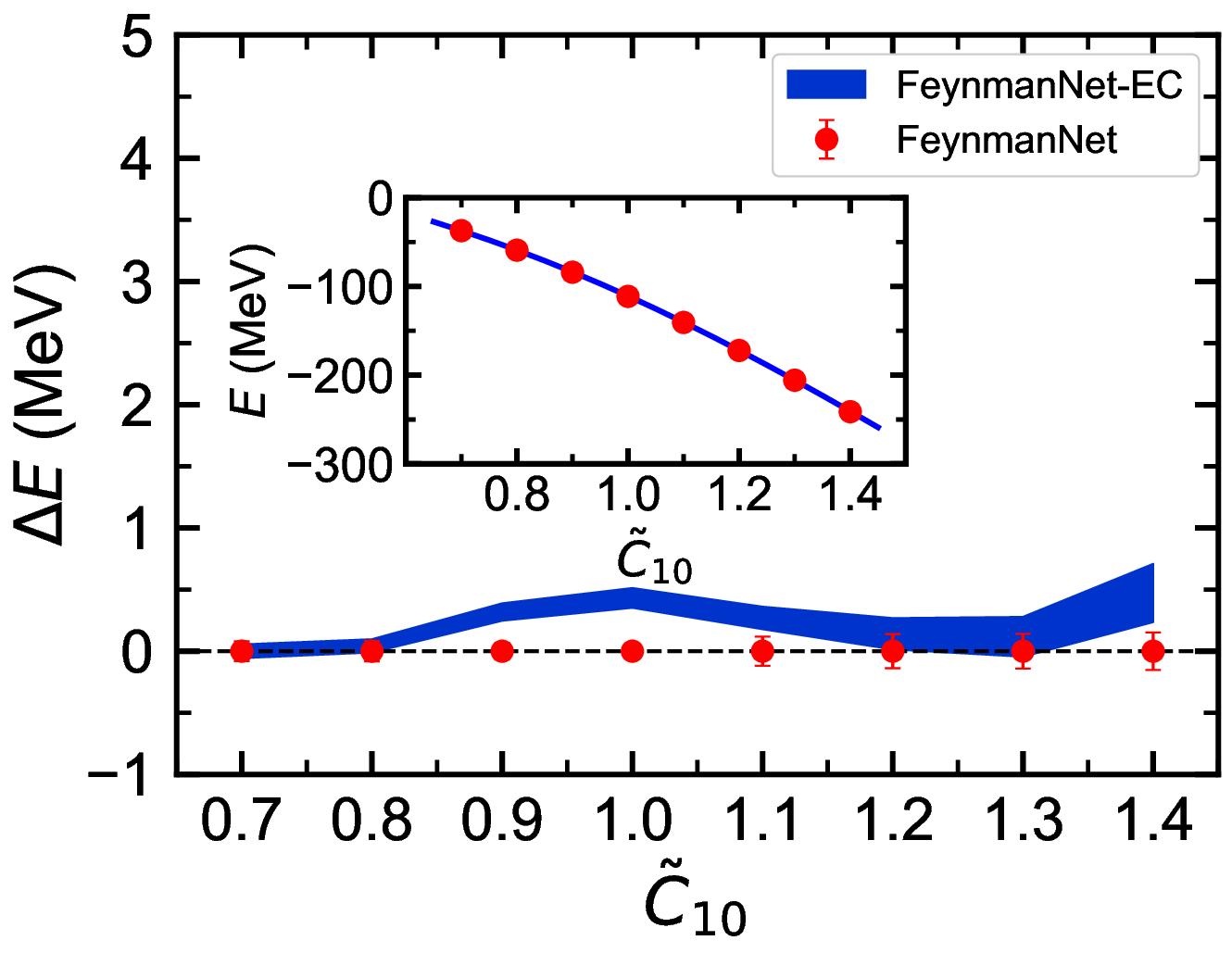}
  \caption{The  differences \(\Delta E =E - E_{\mathrm{ref}}\) from \(E\), the ground-state energies of \({}^{15}\mathrm{O}\) obtained with \emph{FeynmanNet}-EC (blue band) and \emph{FeynmanNet} (red dot), to \(E_{\mathrm{ref}}\), the central values of the \emph{FeynmanNet} results.
  The width of the blue band and the error bars of the red dot indicate the statistical errors of the \emph{FeynmanNet}-EC and \emph{FeynmanNet} results, respectively. $\tilde{C}_{10}$ denotes the relative change of $C_{10}$ to its reference value, while the other LECs are fixed at their reference values. The inset shows the ground-state energies of ${}^{15} \rm O$ from \emph{FeynmanNet}-EC and \emph{FeynmanNet}, as functions of $C_{10}$.}
  \label{fig:bootstrap_ec}
\end{figure}
In Fig.~\ref{fig:bootstrap_ec}, taking ${}^{15} \rm O$ as an example, the \emph{FeynmanNet}-EC emulated ground-state energies are compared to those from direct \emph{FeynmanNet} calculations.
The LEC $C_{10}$ is varied around its reference value by $\pm20\%$, while the other two LECs are fixed at their reference values.
Importantly, these five LEC points are not used as training samples for training the EC emulator (see Fig.~\ref{fig:lec_samples}) and thus, the \emph{FeynmanNet}-EC results are pure predictions.
The \emph{FeynmanNet}-EC predictions are in excellent agreement with direct \emph{FeynmanNet} calculations, in most cases, within statistical uncertainties. 
For each comparison point, the discrepancy between the two calculations is less than $0.5\%$.
Furthermore, the statistical errors are between 0.05 MeV and 0.25 MeV, which are small compared to the variations of ground-state energies induced by LEC changes.
This indicates that statistical noise from the EC procedure should have negligible effects in the present calculations of sensitivity indices.
Consequently, in the following sensitivity analyses, we neglect the statistical errors from \emph{FeynmanNet}-EC emulations.

\begin{figure}[!htbp]
  \centering
  
  \begin{subfigure}{0.9\linewidth}
    \centering
    \includegraphics[width=1\linewidth]{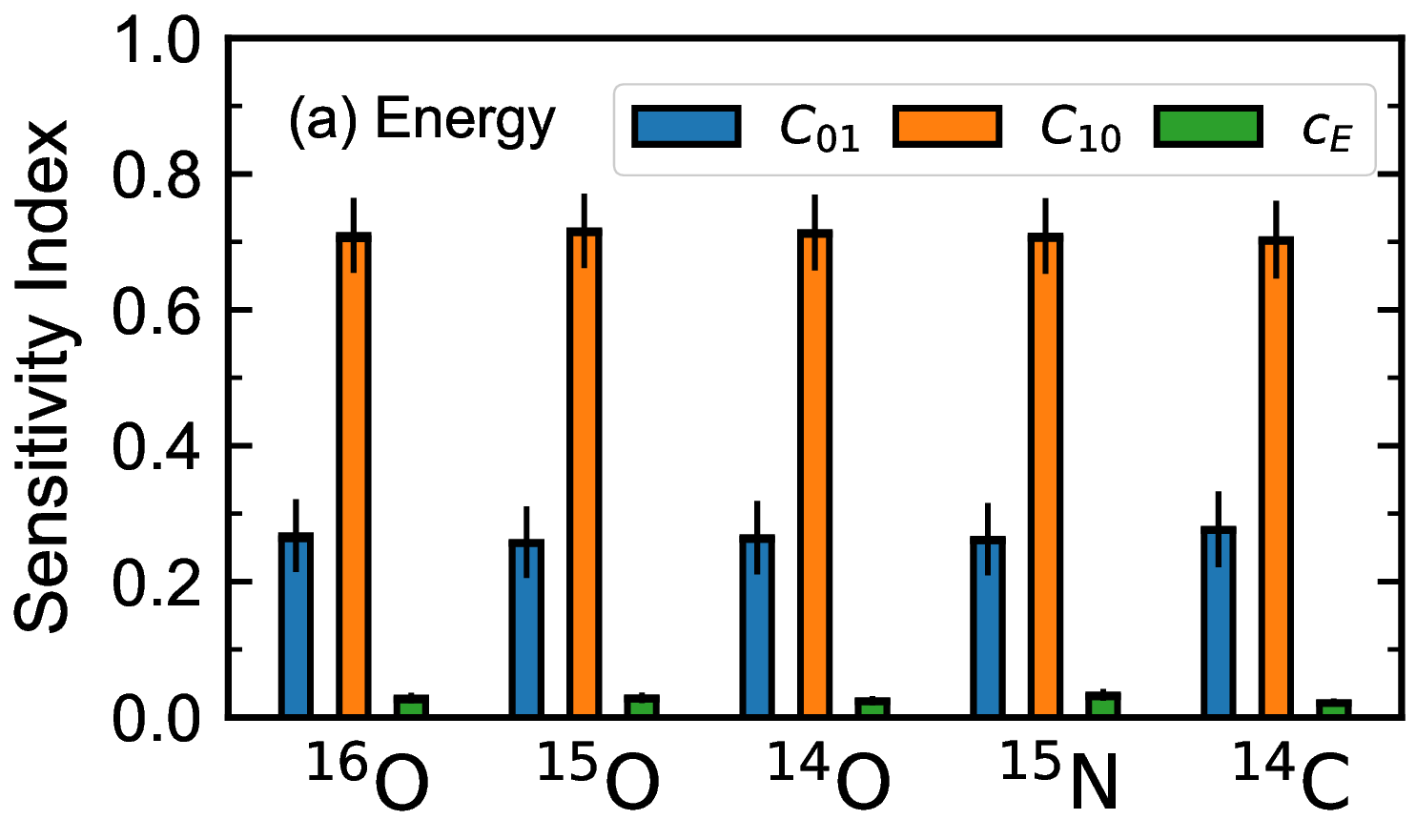}
  \end{subfigure}
  \vspace{0.3cm}
  
  \begin{subfigure}{0.9\linewidth}
    \centering
    \includegraphics[width=1\linewidth]{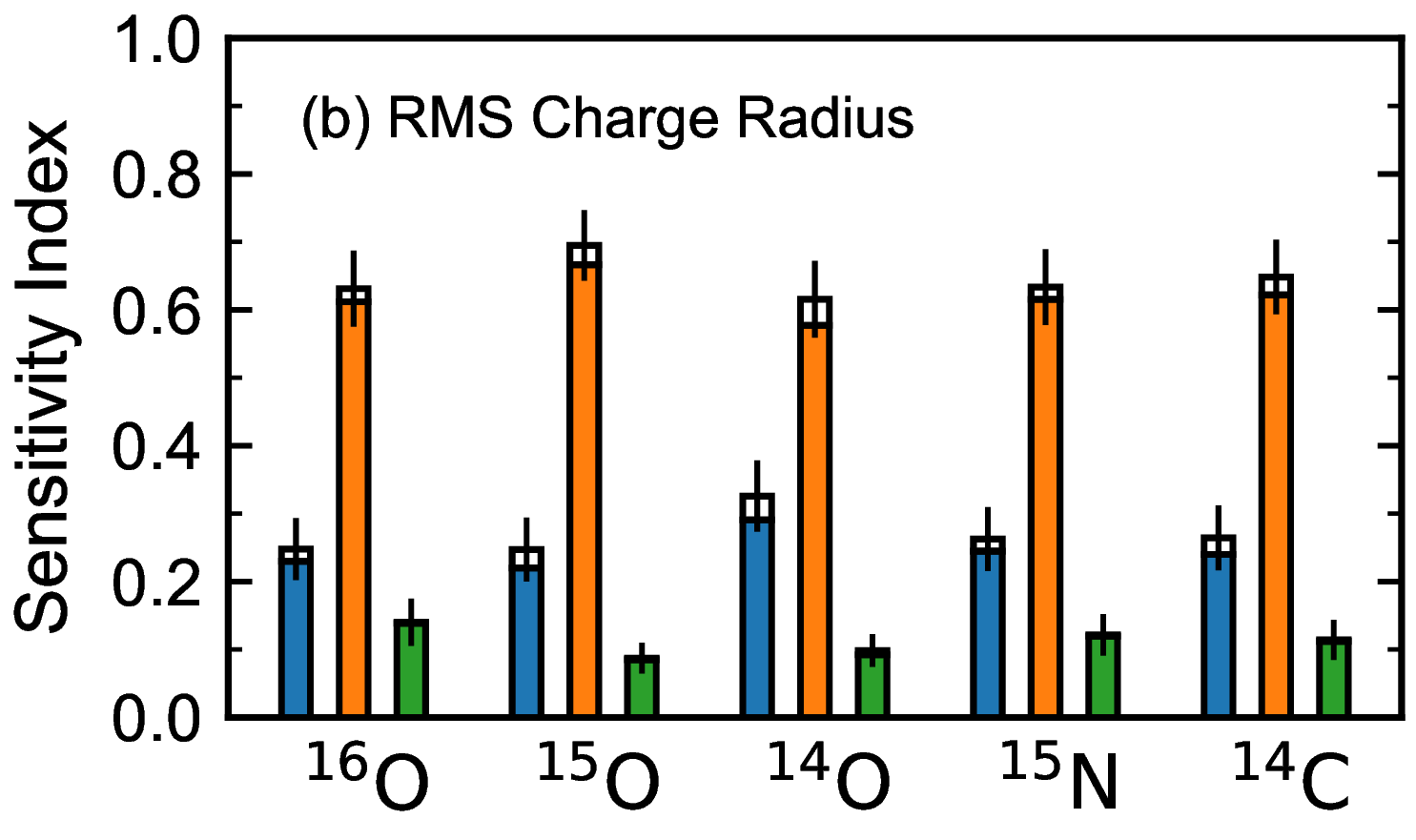}
  \end{subfigure}
  \vspace{0.3cm}
  
  \begin{subfigure}{0.9\linewidth}
    \centering
    \includegraphics[width=1\linewidth]{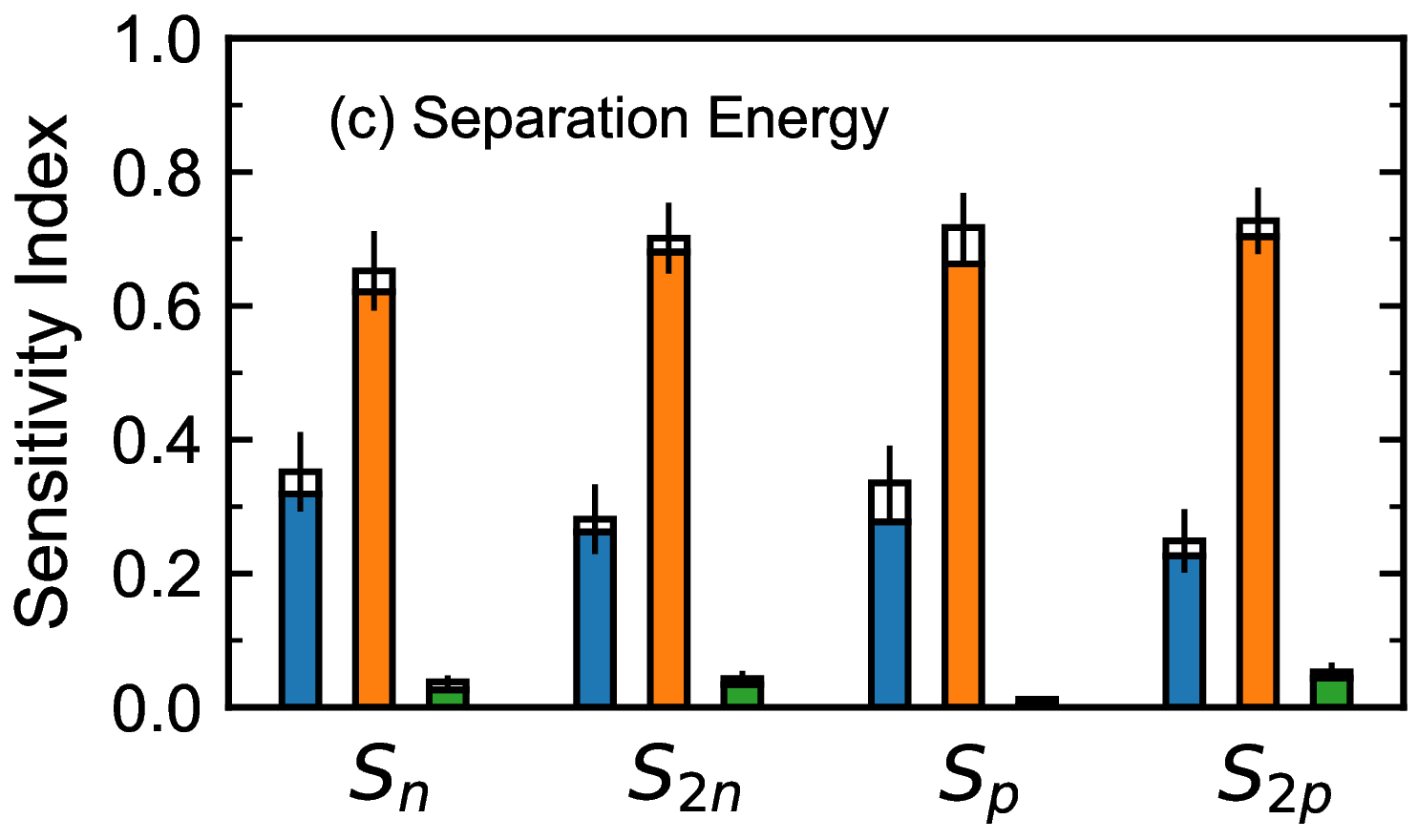}
  \end{subfigure}

  \caption{Main-effect and total-effect Sobol indices for the (a) ground-state energies and (b) rms charge radii of
   \({}^{16}\mathrm{O}, {}^{15}\mathrm{O}, {}^{14}\mathrm{O}, {}^{15}\mathrm{N}\), and \({}^{14}\mathrm{C}\), as well as (c) the one- and two-nucleon separation energies ($S_n, S_{2n}, S_p, S_{2p}$) for \({}^{16}\mathrm{O}\).
   For each nucleus, the three vertical bars represent the Sobol indices of three LECs, respectively. In each bar, the filled portion denotes the main effect, while the full bar height corresponds to the total effect. Error bars denote 68\% confidence interval.}
  \label{fig:sobol_all}
\end{figure}

Figure~\ref{fig:sobol_all} (a) shows the GSA results of the five nuclei, ${}^{16}\mathrm{O}$, ${}^{15}\mathrm{O}$, ${}^{14}\mathrm{O}$, ${}^{15}\mathrm{N}$, and ${}^{14}\mathrm{C}$,
using the essential Hamiltonian derived from LO pionless EFT.
For the three LECs ($C_{01}, C_{10}, c_E$) in the essential Hamiltonian, the main effects are nearly identical to the total effects for all the nuclei considered.
The second-order correlations between the LECs are minimal, about $1.19\%$ for ${}^{16}\mathrm{O}$, $0.64\%$ for ${}^{15}\mathrm{O}$, $0.63\%$ for ${}^{14}\mathrm{O}$, $0.85\%$ for ${}^{15}\mathrm{N}$, and $0.52\%$ for ${}^{14}\mathrm{C}$. 
On average, they account for only $\approx0.77\%$ of the total variance. 
The present results reveal that the energy is almost additive in all LECs of the essential Hamiltonian, and nonlinear interactions in the energy between LECs are weak.

The dominant contribution to the sensitivity arises from $C_{10}$ for all the considered nuclei, 
with the main effect around $0.70$. 
The next most influential parameter is $C_{01}$, with values ranging from $0.26$-$0.28$.
By contrast, the three-body contact term $c_E$ shows very small main and total effects, typically in the range $0.02$-$0.03$.
The above pattern for the GSA of $^{16}$O ground-state energies from the essential Hamiltonian is consistent with that from the realistic chiral EFT Hamiltonian~\cite{Ekstrom2019}.
The present GSA analysis extends this pattern to the ground-state energies of other nearby open-shell nuclei.
This supports the mechanism underlying the success of the essential Hamiltonian proposed in Ref.~\cite{Gnech2024Phys.Rev.Lett.142501}, namely, the low-energy observables are insensitive to the short-range details of the nuclear force.

Figure~\ref{fig:sobol_all} (b) presents the GSA results for the rms charge radii of the same set of nuclei.
In general, the pattern is fairly similar to the energy sensitivity analysis.
The contributions from second-order correlations amount to less than $\approx 2.3\%$ across all five nuclei.
The overall variance decomposition shows that $C_{10}$ is the dominant source of sensitivity, 
with main effects in the range $0.60$-$0.65$, while $C_{01}$ provides the next largest contribution, around $0.22$-$0.27$.
The contribution from $c_E$, however, is somewhat more pronounced than those in the energy sensitivity analysis, typically at the level of $0.09$-$0.14$.
The above pattern for the GSA of rms charge radii from the essential Hamiltonian is also consistent with that obtained from the realistic chiral EFT Hamiltonian~\cite{Ekstrom2019}.
This further corroborates the effectiveness of the essential Hamiltonian.

Figure~\ref{fig:sobol_all} (c) presents the GSA results for the one- and two-nucleon separation energies $S_n$, $S_{2n}$, $S_p$, and $S_{2p}$ for $^{16}$O.
For the one- and two-nucleon separation energies, the overall picture still resembles that of the rms charge radii: the variance is dominated by $C_{10}$, with contributions around $0.62$-$0.70$, while $C_{01}$ accounts for $0.23$-$0.32$, and $c_E$ remains small ($<0.05$ in most cases).
The second-order correlations are also small, similar to those in the sensitivity analysis of rms charge radii.
This indicates that the one- and two-nucleon separation energies are also insensitive to the short-range details of the nuclear force.

It should be noted that the present GSA study primarily characterizes the behavior of the essential Hamiltonian for bulk properties. Whether these sensitivity patterns are similar for high-resolution Hamiltonians or for structure phenomena beyond S-wave-dominated properties remains to be investigated.

\begin{figure}[!htbp]
  \centering
  \includegraphics[width=1\linewidth]{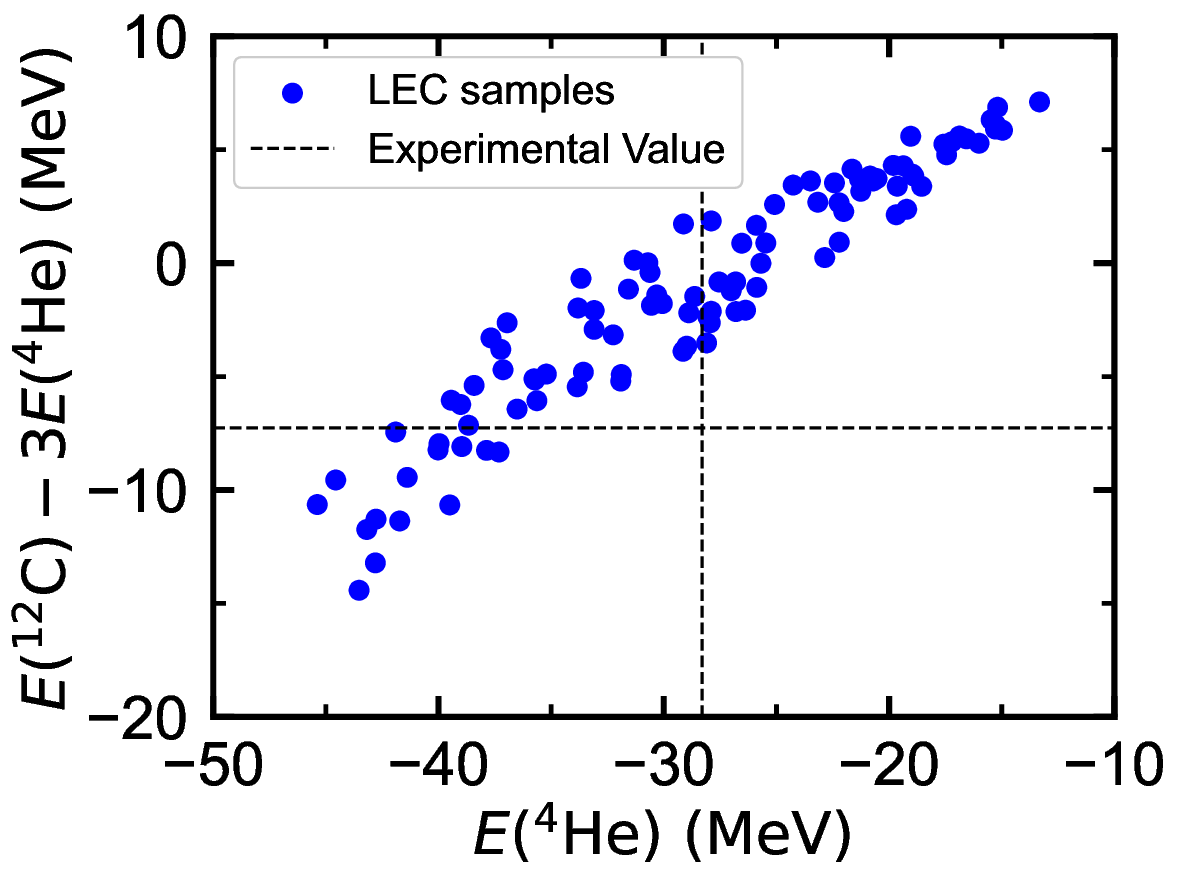}\\[2ex]
  \includegraphics[width=1\linewidth]{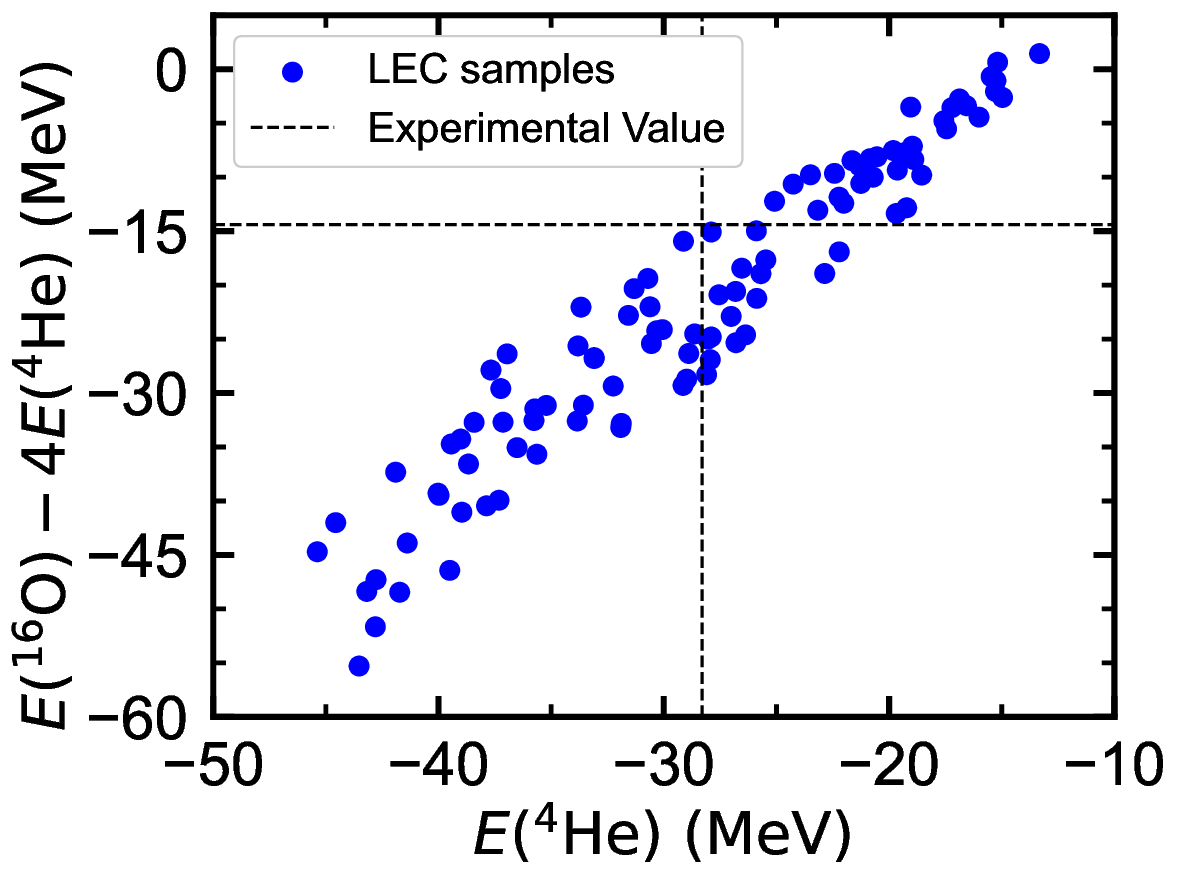}
  \caption{The ground-state energies of $^{12}$C (top) and $^{16}$O (bottom) with respect to the multialpha breakup threshold, $E(^AZ)-E(^4{\rm He})A/4$, obtained from the essential Hamiltonian under random variations of LECs, as functions of the alpha-paticle energy. 
  The black dashed lines indicate the corresponding experimental values.}
  \label{fig:He4_C12_O16_correlations}
\end{figure}
Finally, we investigate the description of alpha-alpha interactions from the essential Hamiltonian, whose strength can be characterized by the alphalike nuclear ground-state energies with respect to the multialpha breakup threshold, $E(^AZ)-E(^4{\rm He})A/4$.
Figure \ref{fig:He4_C12_O16_correlations} depicts the correlation between $E({}^{4}\mathrm{He})$ and $E(^AZ)-E(^4{\rm He})A/4$ for $^{12}$C and $^{16}$O.
For both ${}^{12}\mathrm{C}$ and ${}^{16}\mathrm{O}$, a robust correlation exists between $E({}^{4}\mathrm{He})$ and $E(^AZ)-E(^4{\rm He})A/4$.
This suggests that the strength of alpha-alpha interactions is correlated with the interactions between nucleons.

However, the $E(^AZ)-E(^4{\rm He})A/4$ values from the essential Hamiltonian with varying LECs miss the experimental values.
In particular, when the alpha-particle energy is reproduced,  the theoretical $E({}^{12}\mathrm{C})-3E({}^{4}\mathrm{He})$ values are higher than the experimental value, while the theoretical $E({}^{16}\mathrm{O})-4E({}^{4}\mathrm{He})$ values are systematically lower than the experimental value.
Such discrepancy cannot be compensated by the variation of LECs due to the strong correlation between $E(^AZ)-E(^4{\rm He})A/4$ and the alpha-particle energy.
It is possible to modify this correlation by changing the range $R_3$ of the repulsive 3N interactions [Eq.~\eqref{eq:3NF}].
By increasing $R_3$ and adjusting the 3N LEC $c_E$ simultaneously, one could produce both the correct $^4$He and $^{16}$O energies~\cite{Gnech2024Phys.Rev.Lett.142501}.
However, varying $R_3$ cannot solve the deviations in $^{12}$C and $^{16}$O simultaneously, because the theoretical results of two nuclei deviate from the experimental values in opposite directions, as shown in Fig.~\ref{fig:He4_C12_O16_correlations}.

The present results indicate that there are still deficiencies in describing the alpha-alpha interactions from the essential Hamiltonian~\cite{Schiavilla2021}.
Two factors may underlie these deficiencies.
First, the essential Hamiltonian is purely local, while the lattice nuclear EFT studies reveal a strong dependence of alpha-alpha interactions on the degree of locality of NN interactions~\cite{Elhatisari2016}.
Second, spin-orbit interactions are missing in the essential Hamiltonian.
The importance of the spin-orbit interactions was highlighted in the recent study~\cite{Niu2025}, in which a LO SU(4) Hamiltonian augmented by spin-orbit interactions is able to describe well the binding energies from $^4$He to $^{132}$Sn, as well as nuclear saturation.
In the future, it should be crucial to improve the essential Hamiltonian in these two aspects to extend its accurate predictions to heavier nuclei.

In general, a 0.5\% emulation error of \emph{FeynmanNet}-EC translates to an absolute energy scale that is large enough to blur delicate alpha-cluster thresholds (see Fig.~\ref{fig:bootstrap_ec}). However, as demonstrated in Fig.~\ref{fig:He4_C12_O16_correlations}, the deviations between the predictions of the essential Hamiltonian and the experimental values for $^{12}$C and $^{16}$O are much larger. Therefore, the emulation error does not threaten the above qualitative physical conclusions regarding the alpha-alpha interactions.

\section{Summary and outlook}
We have developed an efficient emulator for \emph{ab initio} calculations of nuclear ground-state properties, named \emph{FeynmanNet}-EC, by combining the \emph{FeynmanNet} framework of neural-network variational Monte Carlo and the eigenvector continuation method.
We apply the \emph{FeynmanNet}-EC emulator to compute the ground states of several selected nuclei, ${}^{16}\mathrm{O}$, ${}^{15}\mathrm{O}$, ${}^{14}\mathrm{O}$, ${}^{15}\mathrm{N}$, and ${}^{14}\mathrm{C}$, as emerged from an essential Hamiltonian ``o" inspired by leading-order pionless EFT~\cite{Schiavilla2021}.
With a small number of EC samples $N_{\rm EC}=8$, the \emph{FeynmanNet}-EC emulator can reach a $0.5\%$ accuracy compared to the full \emph{FeynmanNet} solutions of the binding energies.
Then, we perform a global sensitivity analysis of the binding energies, separation energies, and rms charge radii of the selected $A\simeq 16$ nuclei for the three LECs in the essential Hamiltonian.
From the global sensitivity analysis, it is found that the variance of these bulk properties is almost additive in the three LECs, as the main effects closely match the total effects, and the second-order correlations are negligible. 
The bulk properties are most sensitive to the two-body LEC $C_{10}$ in the $^3S_1$ channel and insensitive to the LEC $c_E$ of the three-body short-range interaction.
Next, we investigate the correlation among $^4$He, $^{12}$C, and $^{16}$O binding energies by varying the LECs in the essential Hamiltonian.
The analysis reveals robust correlations between both $^{12}$C-$^4$He and $^{16}$O-$^4$He binding energies.
Nevertheless, the experimental ground-state energies of $^{12}$C and $^{16}$O cannot be reproduced simultaneously by varying the LECs
in the Hamiltonian. 
The discrepancy may indicate that additional ingredients are
required to improve the leading order Hamiltonian for its description of light nuclei.

The \emph{FeynmanNet}-EC emulator provides an efficient tool for identifying how each LEC contributes to the variances of the ground-state energies, charge radii and separation energies.
In the future, it will be straightforward to apply the \emph{FeynmanNet}-EC emulator to heavier nuclei, thanks to its computational efficiency and favorable scaling with $A$.
\begin{acknowledgments}
ML thanks Tianxing Huang for helpful discussion at various stages of this work. 
This work has been supported in part by the National Key R\&D Program of China (Contracts No. 2024YFA1612600, No. 2024YFE0109803), 
the National Natural Science Foundation of China (Grants No. 12475117, No. 12141501, and No. 12435006), 
Beijing Natural Science Foundation (Grant No. QY24009), the High-performance Computing Platform of Peking University,
and the National Key Laboratory of Neutron Science and Technology (Grant No. NST202401016).
\end{acknowledgments}

\end{document}